\documentstyle[12pt]{article}
\begin{document}

\baselineskip 20pt 

\vspace{.5in}

\begin{center}

{\Large \bf   
Relativistic corrections to\\the energy spectra of\\completely confined 
particles\\
}
\vskip .8in
{\bf
Shang Yuan Ren\\
Department of Physics, Peking University \\
Beijing 100871, People's Republic of China\\
}
\vbox{}
\end{center}
\newpage
\begin{center}
{\bf \large Abstract}
\end{center}
\par
An analytical expression for the relativistic corrections to the energy spectra
of particles completely confined in an one-dimensional limited length in real
space is given, based upon the wave property of particles, the relativistic 
energy-momentum relation and two mathematical equations. 
\newpage
\par
The quantum confinement is one of the most fundamental problems in low
dimensional physics. The expression for the energy spectra of a non-relativistic
particle of mass $m$ confined in an one-dimensional limited length
in real space between $ x = -L/2 $ and $ x = L/2 $
has been given in almost any standard quantum mechanics
textbook as a classical example of solving the one
dimensional Schr$\ddot{\rm o}$dinger differential equation with 
infinite potential barriers and is well known as
\begin{equation}
E_j = \frac { j^2 {\hbar}^2 {\pi}^2}{2 m L^2},
\end{equation}
where j is a postive integer.
This result is widely used in condensed matter physics, as a theoretical basis
of many potential applications of low-dimensional quantum confinement
devices\cite{rvs}.\par
However, as the energy of the confined particle increases, the relativistic
effect will show up and the equation (1), which was based on the solution
of the non-relativistic Schr$\ddot{\rm o}$dinger differential equation, 
must be modified. As the energy
further increases to very large then there could be even creations of
new particles. Here we are interested in the energy range of the particle that
the relativistic effect could show up but no new particle's creations.\par
The usual way of obtaining (1) - solving the Schr$\ddot{\rm o}$dinger 
differential equation with infinite potential barriers - is not easy to extend
to the relativistic case straightforwardly\cite{lan}.
In the following we use a different
approach to give the result of equation (1). The corresponding solution with
the relativistic effect can be obtained naturally.
\par
We look the problem from a different way:
All plane waves 
\begin{equation}
 \phi(k,x) = \sqrt{ \frac{ 1 }{ 2 \pi } }e^{ikx} ~~~~~~~~~-\infty~<~k~<~+\infty
\end{equation}
are a complete set $C$,
from this complete set $C$ we can always construct a
subset of wavefunctions $IN \subset C$ by requiring
that every wavefunction $\psi_j(x) \in IN $ to be
a standing wave inside the confined real space $-L/2 < x < L/2$ but 
zero everywhere outside this region.
The standing wave of $j-1$ nodes is
\begin{eqnarray*}
\psi_j(x) = C_j sin(\frac{j\pi}{L}(x-\frac{L}{2})), ~~~~~~if~~ | x |~ < \frac{L}{2}, 
\end{eqnarray*}
\begin{equation}
~~~~~~~~ = 0, ~~~~~~~~~~if~~ | x |~ \geq \frac{L}{2}.~~~~~~~~~~~~~~~~~~~~~~~~~ 
\end{equation}
That is, we could choose combination coefficients $c_j(k)$ to get
\begin{equation}
\psi_j(x) = \int c_j(k) \phi(k,x) dk,
\end{equation}
and to require $\psi_j(x)$ to satisy the equation (3).\par
Of course,
\begin{equation}
c_j(k) = \frac{ j \sqrt { \pi L}}{ j^2 {\pi}^2 - k^2 L^2 } 
( e^{-ik \frac{L}{2}} - (-1)^j e^{ik \frac{L}{2}} ) 
\end{equation}
is the normalized Fourier transform of $\psi_j(x)$.
\par
The probability that the confined state $\psi_j(x)$ has wavevector from $k$ to
 $k+dk$ is $|c_j(k)|^2dk $.
Concerning $|c_j(k)|^2$, there are two mathematical equations\cite{tab}
\begin{equation}
\int_{-\infty}^{\infty}  
\frac{4 j^2 \pi L}{( j^2 {\pi}^2 - k^2 L^2)^2} cos^2(kL/2) dk = 1
\end{equation}
and 
\begin{equation}
\int_{-\infty}^{\infty} 
\frac{4 j^2 \pi L}{( j^2 {\pi}^2 - k^2 L^2)^2} cos^2(kL/2) k^2 dk  = 
\frac { j^2 {\pi}^2 } {L^2}
\end{equation}
for odd number $j$ and similar equations but with the sine function instead of
the cosine function for even number $j$. 
\par
If the plane wave $\phi(k,x)$ corresponds an eigen value $ o(k)$ of
an operator {\bf O}, which is a function of $k$, then the corresponding 
expectation values of the operator {\bf O} in the confined state
$\psi_j(x)$ can be obtained as,
\begin{equation}
O_j = \int |c_j(k)|^2 o(k) dk.
\end{equation}
It may be noticed that so far {\it only the wave property of particles is
 used}.\par 
Based upon equations (6)-(8), we can discuss several different cases:\\
For non-relativistic particles with a mass $m$, the operator {\bf O} we are
interested is simply the energy operator {\bf E}. The energy momentum relation
of free non-relativistic particles is 
\begin{equation}
\varepsilon (k) = \frac { {\hbar}^2 k^2}{2m},
\end{equation}
naturally
\begin{eqnarray*}
E_j = \int |c_j(k)|^2  \frac { {\hbar}^2 k^2}{2m} dk
\end{eqnarray*}
\begin{equation}
~~~~ = \frac { j^2 {\hbar}^2 {\pi}^2}{2 m L^2}
\end{equation}
by using equation (6) and (7).
This is exactly the same as equation (1).\\
For particles without mass, we have {\bf O} = {\bf $\Omega^2$} and 
\begin{equation}
 {\omega}^2(k) = c^2 k^2,
\end{equation}
therefore
\begin{equation}
E_j = \hbar \Omega_j = \hbar c \frac{j \pi}{L}.
\end{equation}
For relativistic particles with static mass $m$, we have {\bf O} = {\bf E$^2$}
and the corresponding dispersion relation for free relativistic particles is 
\begin{equation}
\varepsilon^2(k) = m^2 c^4 + {\hbar}^2 c^2 k^2,
\end{equation}
therefore we have 
\begin{equation}
E^2_j = m^2 c^4 + j^2 \hbar^2 c^2 (\frac{ \pi }{ L })^2
\end{equation}
by using (6) and (7). Or 
\begin{equation}
E_j = \sqrt{ m^2 c^4 + j^2 \hbar^2 c^2 (\frac{ \pi }{ L })^2 }.
\end{equation}
In the non-relativistic limit the equation (15) return to equation (10), in
addition to a static energy term $m c^2$. The author is not aware of (14) or
(15) having been published before.\par
The extension of (14) or (15) to the 
two or three-dimensional confined relativistic particles is straightforward.
\vskip .12in
Acknowledgment:
\par
The author is grateful to Professors Kun Huang, John D. Dow, L. Z. Fang,
C. S. Gao, H. Y. Guo, J. Y. Zeng and G. D. Zhao for stimulating discussions. 
This research is supported by the National Natural Science Foundation of China 
(Climbing program on Science of Nano-materials and Projects No. 19574008 and No.
19774001). \\

\end{document}